\documentclass[slac_one]{revtex4}
\usepackage{graphicx}
\usepackage{fancyhdr}
\usepackage{xspace}
\newcommand{\ttbar}{\ensuremath{t \bar t}\xspace}

\newcommand{\Wj    }{\ensuremath{\Wboson\mbox{+jets}}\xspace}

\newcommand{\btag   }{\ensuremath{b\mbox{-tagging}}\xspace}

\newcommand{\bquark }{\ensuremath{b\mbox{-quark}}\xspace}
\newcommand{\bquarks}{\ensuremath{b\mbox{-quarks}}\xspace}

\newcommand{\ifb}{\mbox{fb$^{-1}$}}%  Inverse femtobarns.
\newcommand{\inb}{\mbox{nb$^{-1}$}}%  Inverse nanobarns.
\newcommand{\Wboson}{\ensuremath{W}}%
\newcommand{\Zboson}{\ensuremath{Z}}%
\newcommand{\TeV}{\ifmmode {\mathrm{\ Te\kern -0.1em V}}\else
               \textrm{Te\kern -0.1em V}\fi}%
\newcommand{\GeV}{\ifmmode {\mathrm{\ Ge\kern -0.1em V}}\else
               \textrm{Ge\kern -0.1em V}\fi}%
\newcommand{\MeV}{\ifmmode {\mathrm{\ Me\kern -0.1em V}}\else
               \textrm{Me\kern -0.1em V}\fi}%
\newcommand{\keV}{\ifmmode {\mathrm{\ ke\kern -0.1em V}}\else
               \textrm{ke\kern -0.1em V}\fi}%
\newcommand{\eV}{\ifmmode  {\mathrm{\ e\kern -0.1em V}}\else
               \textrm{e\kern -0.1em V}\fi}%

\newcommand{\rts} {\ensuremath{\sqrt{s}}}
\newcommand{\pt} {\ensuremath{p_{T}}}
\newcommand{\met} {\ensuremath{E_{\mathrm{T}}^{\mathrm{miss}}}} % Sub/superscript roman not italic (EE)

\newcommand{\lint}{\mbox{$(295 \pm 32)~\inb$}}

\begin{document}

\title{First results on top-quarks from ATLAS} %% Paper title goes here


\author{G. Cortiana, on behalf of the ATLAS Collaboration}
\affiliation{Max-Planck-Institut f\"{u}r Physik, F\"{o}hringer Ring 6, D-80805 Munich, Germany}

\begin{abstract}

The search for first \ttbar candidate events and the related
background studies using data-driven techniques are
reported for about 300~\inb\ of \rts=7~\TeV\ proton-proton collision data
delivered by the Large Hadron Collider (LHC) and collected with the
ATLAS detector. Selected events are characterized by the presence of
high-\pt\ isolated charged lepton(s), high jet multiplicity, jet(s)
identified as originating from \bquark\ by a secondary vertex tagger
algorithm, and missing transverse energy. They reveal kinematics
properties consistent with top pair production.

\end{abstract}

\maketitle

\thispagestyle{fancy}


\section{Introduction} % Section title should be in all capitals.

During the operation at the center-of-mass energy
(\rts) of 7~\TeV, the LHC is expected to deliver up
to 1~\ifb\  of $pp$ collision data by the end of 2011.
The top pair production cross section at this energy is expected to be
about 160~pb \cite{ref:xsec}, approximately 20 times the corresponding
production cross section at the Tevatron collider. 
Already at this initial stage of the data taking with the ATLAS 
experiment~\cite{ref:atlas}, and with very small data
samples, corresponding to an integrated luminosity of \lint, top pair
candidate events are searched for, and preliminary data-driven
background studies are carried out. After a short presentation of the key
ingredients to top-quark physics analysis, provided in
Section \ref{sec:ingredients}, early results are discussed in
Section~\ref{sec:cand} and Section~\ref{sec:bkg}.

\section{Ingredients to top-physics: datasets and object selection}
\label{sec:ingredients}

Top pair final states are classified according to the \Wboson\ boson
decays. The all-jet mode accounts for about 46\% of the decays, and
lepton plus jets and dilepton modes for about 44\% and 10\% of the
decays respectively.  Final states containing electrons or muons are
of particular interest for early measurements as they provide clear
trigger signals and rich event signatures.
The events contain jets (two of which originate from \bquarks),
high \pt, isolated charged lepton(s), and missing transverse energy,
\met, from the escaping neutrino, and explore the complete detector
capabilities.

During the initial data taking period only the first of the three
level trigger architecture functionalities available in ATLAS have
been exploited, allowing for the commissioning of the higher level
trigger algorithms and infrastructure. The datasets used for the
analysis presented in this paper correspond to an integrated
luminosity of \lint, and have been collected with electron or muon
triggers, requiring localized energy deposits in the electromagnetic
(EM) calorimeters exceeding a $10\GeV$ threshold, or hit patterns in
the muon spectrometer consistent with muons with $\pt>10\GeV$ originating
from the interaction point.  In addition, early data collected using
minimum bias triggers, requiring coincidence with bunch
crossing of scintillators signals at both detector sides, and jet based
triggers, have been used to study and validate physics objects
identification and reconstruction recipes, and the corresponding
Monte Carlo descriptions~\cite{ref:ele, ref:muo, ref:W, ref:Z,
ref:jets, ref:jes, ref:met, ref:btag}.

Electron candidates are required to fulfill the medium electron
definition~\cite{ref:ele}, which in addition to minimal track quality
and hits requirements and electromagnetic shower shape
information, adopts cluster-to-track matching criteria. Electron
candidates must have $\pt > 20\GeV$, and be within the good detector
acceptance ($|\eta|<2.47$, excluding the calorimeter transition region
$1.37 <|\eta| < 1.52$). To remove photon conversions, the
corresponding track must have an associated hit in the innermost pixel
layer (b-layer hit requirement). To reduce the jet mis-identification
rate and contributions from heavy flavor decays inside jets, the
candidates are required to be isolated: the energy deposition in the
calorimeter within a cone of radius $R
=\sqrt{\Delta\eta^2+\Delta\phi^2}= 0.2$ must be less than $4\GeV +
0.023\cdot \pt^{\rm{ele}}$.

Muons are reconstructed by combining tracks from the inner detector
and the muon spectrometer as defined in~\cite{ref:muo}.  Candidate
muons are required to have $\pt > 20\GeV$ and $|\eta| < 2.5$. To
ensure isolation, the energy deposition in the calorimeter, and the
sum of track transverse momenta measured in a cone of radius $R = 0.3$
around the muon track, are each required to be less than
$4\GeV$. Additionally, the minimum separation between muons and
selected jets is required to be $R=0.4$.

Electron and muon reconstruction and identification procedures have
been successfully exploited for the first ATLAS measurements of the
\Wboson\ and \Zboson\ production cross sections at the LHC reported
in~\cite{ref:W, ref:Z}.

Jets are reconstructed using the anti-$k_T$ algorithm with $R$
parameter of 0.4, combining topological clusters in the
calorimeters. The latter are obtained as three-dimensional groups of
noise-suppressed calorimeter cells, meant to follow the shower
development.  To avoid double-counting, jets overlapping with selected
electrons within $\Delta R = 0.2$ are vetoed.  Due to the
non-compensating nature of the ATLAS calorimeter systems, energy scale
corrections are needed.  Within the initial baseline jet correction
scheme, jets are calibrated to the hadronic energy scale using \pt\
and $\eta$ dependent correction factors obtained from
simulation~\cite{ref:jets}.
%In general the agreement of the correction
%factors between data and simulation is at the level of few
%\%.
The associated jet energy scale (JES) uncertainty has been evaluated
using Monte Carlo events simulated with different detector
configurations and hadronic shower models, and comparing the relative
data to Monte Carlo response in various detector
regions~\cite{ref:jes}.  The total relative JES uncertainty ranges
from 8\% (9\%) in the barrel (end-cap) region for jet $\pt< 60\GeV$,
to 6\% (7\%) for higher \pt\ jets.  The main contributions to the
uncertainties originate from imperfect dead material knowledge (5\%),
from shower modeling and noise description (3-4)\%, and from the
absolute energy scale of the calorimeter (3\%).
%In addition, analysis dependent corrections,
%according to the event topology and the jet flavor, and pile-up levels
%needs to be accounted as well.

The \met~\cite{ref:met}, complementing the lepton identification in
selecting genuine $W\to l\nu$ decays, is reconstructed from the vector
sum of all calorimeter cells, resolved into the transverse
plane. Cells not associated to a jet or electron are included at the
electromagnetic scale, {\em i.e.} without correcting for
non-compensating calorimeter effects. Cells associated with jets are
subject to the calibration scheme described above.  Finally,
additional
\met\ refinements based on information from reconstructed leptons ($e/\mu$)
are applied.

The identification of jets originating from \bquarks constitutes an
effective handle to reject backgrounds to \ttbar\ events.
% which is
%explored by means of different algorithms categories. The first type
%of b-tagging algorithms makes use of b-quark relative long life-time,
%and exploits the presence within jets of displaced tracks with respect
%to the primary event vertex. The second categories of taggers is based
%on explicit reconstruction of secondary vertices within jets, and use
%their properties to discriminates between b-, c-, and light jets.  The
%last type of taggers rests on the identification of soft leptons from
%semileptonic b/c decays within jets. Albeit being limited by the
%semileptonic \bquark branching ratio, these algorithm can complement
%life-time or secondary vertex taggers, and can be used for assessing
%b-tagging efficiencies.
%
Currently, the default \btag algorithm is based on the explicit
reconstruction of secondary vertices within jets, using tracks
displaced with respect to the primary vertex. The tag is assigned
using the secondary vertex decay length significance,
$L/\sigma(L)$, corresponding to a b-jet identification efficiency of
50\%, as evaluated from simulated \ttbar\ Monte Carlo
events~\cite{ref:btag}.

%Secondary vertices
%consistent with $K_S$, $\Lambda_0$ decays, conversions, material
%interactions are removed~\cite{ref:btag}.

\section{Search for first top-quark pair candidate events}
\label{sec:cand}

\begin{figure*}[t]
\centering{
\includegraphics[width=0.39\textwidth]{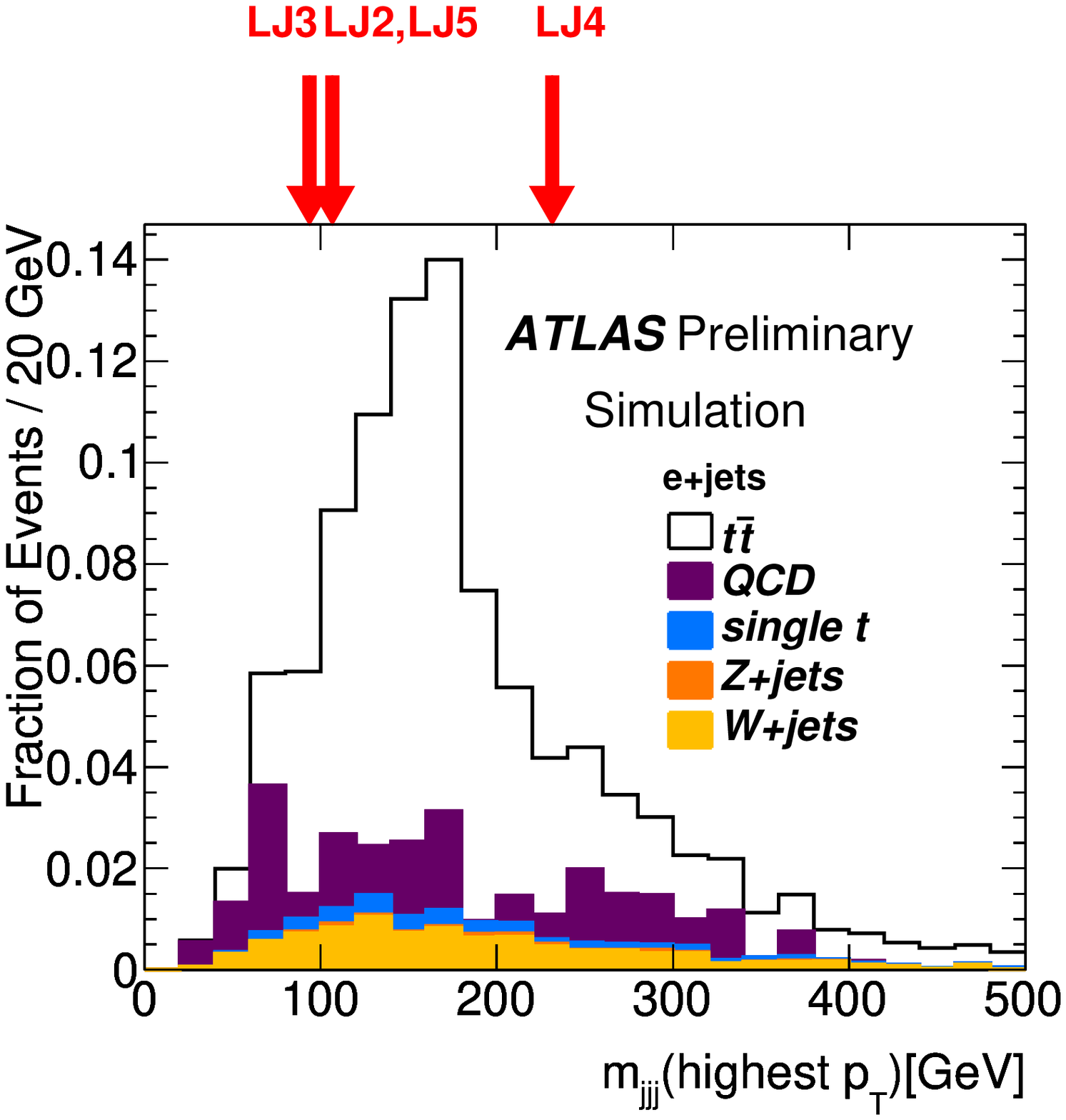}
\includegraphics[width=0.39\textwidth]{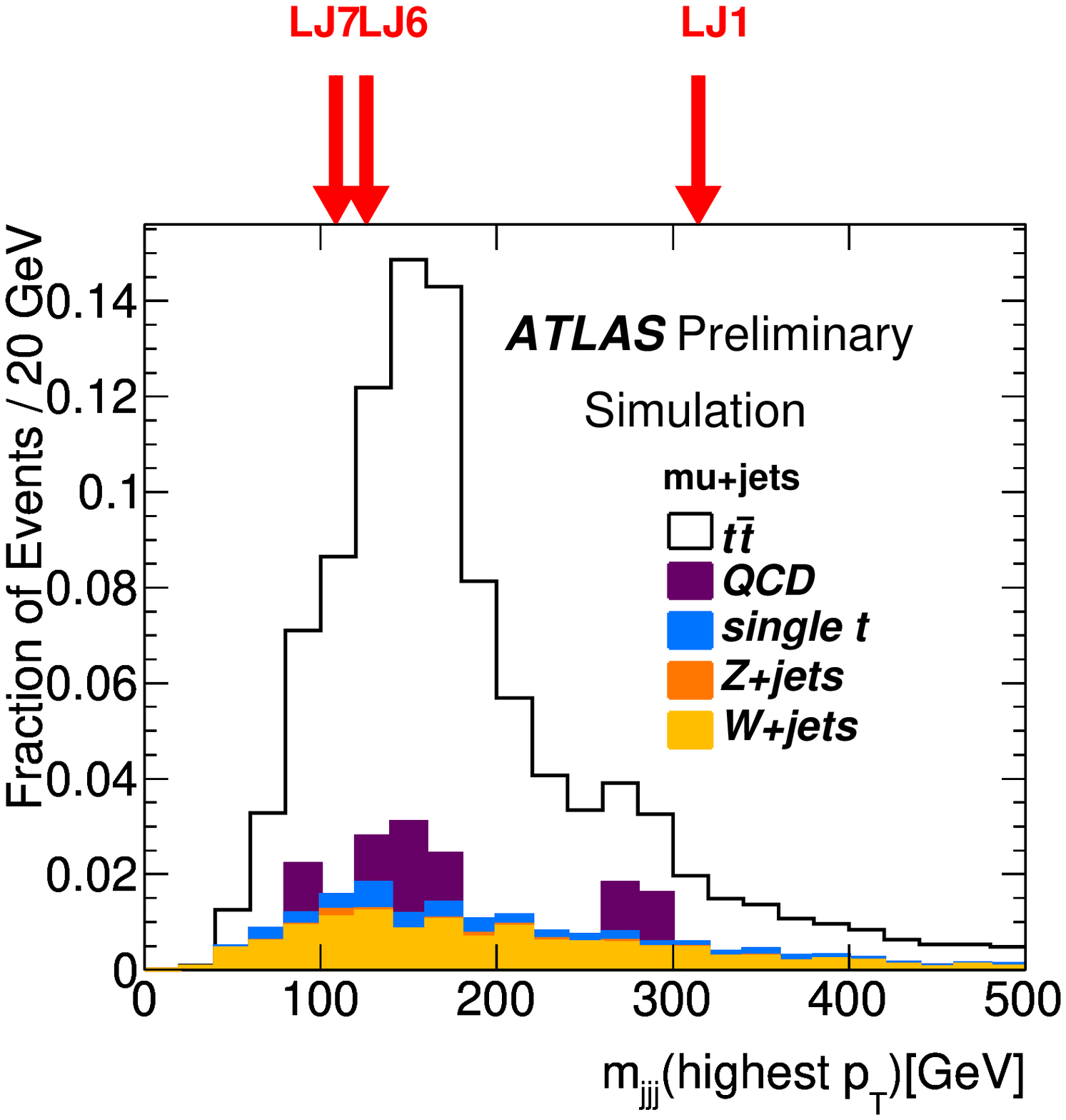}
}
\caption{Invariant mass of the 3-jet combination having the highest \pt\ 
for events passing the electron (left) and muon (right) plus jets \ttbar\ selection.} \label{f:mjjj}

%\end{figure*}
%\begin{figure*}[t]
\centering{
\includegraphics[width=0.34\textwidth]{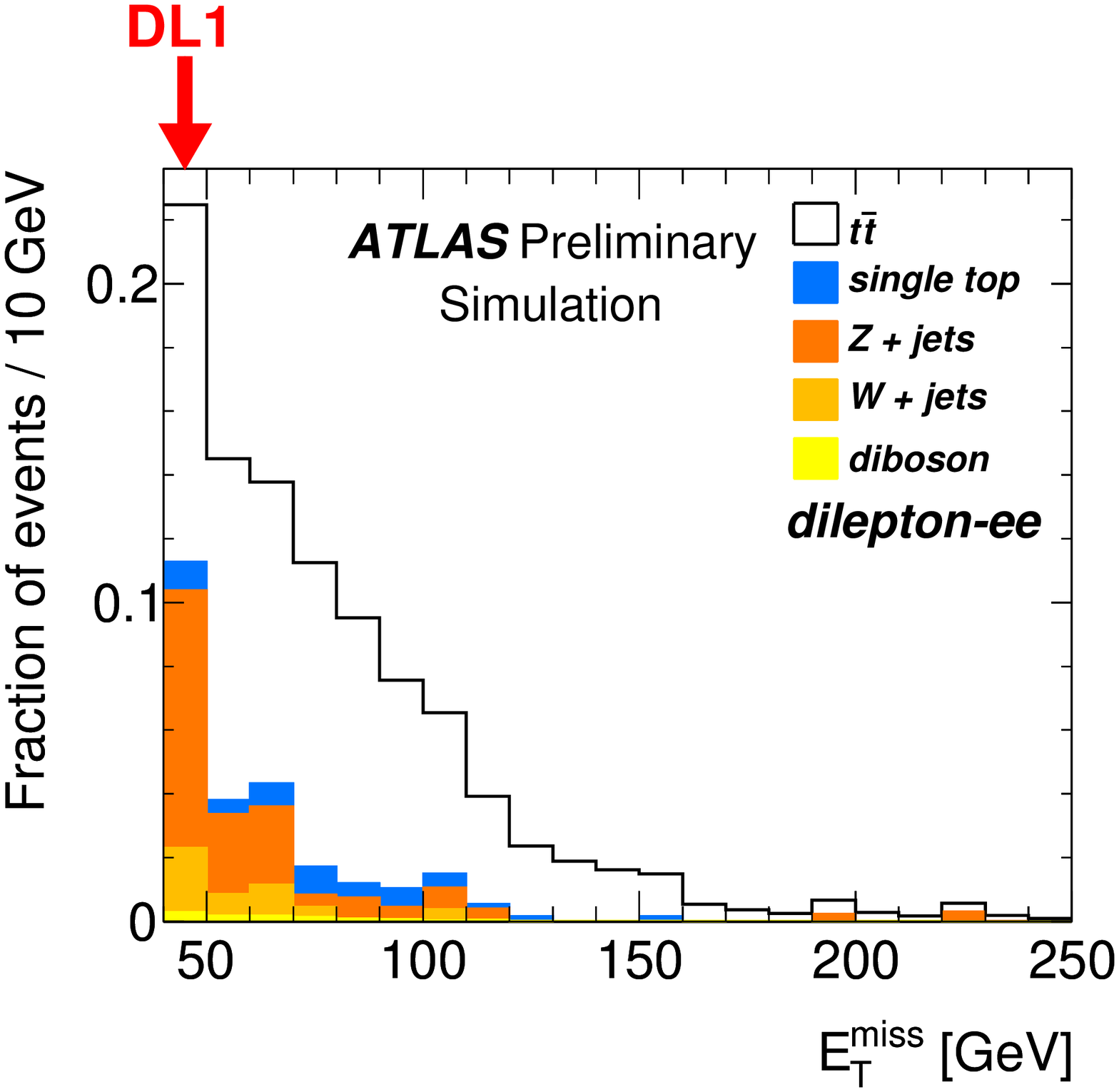}
\includegraphics[width=0.34\textwidth]{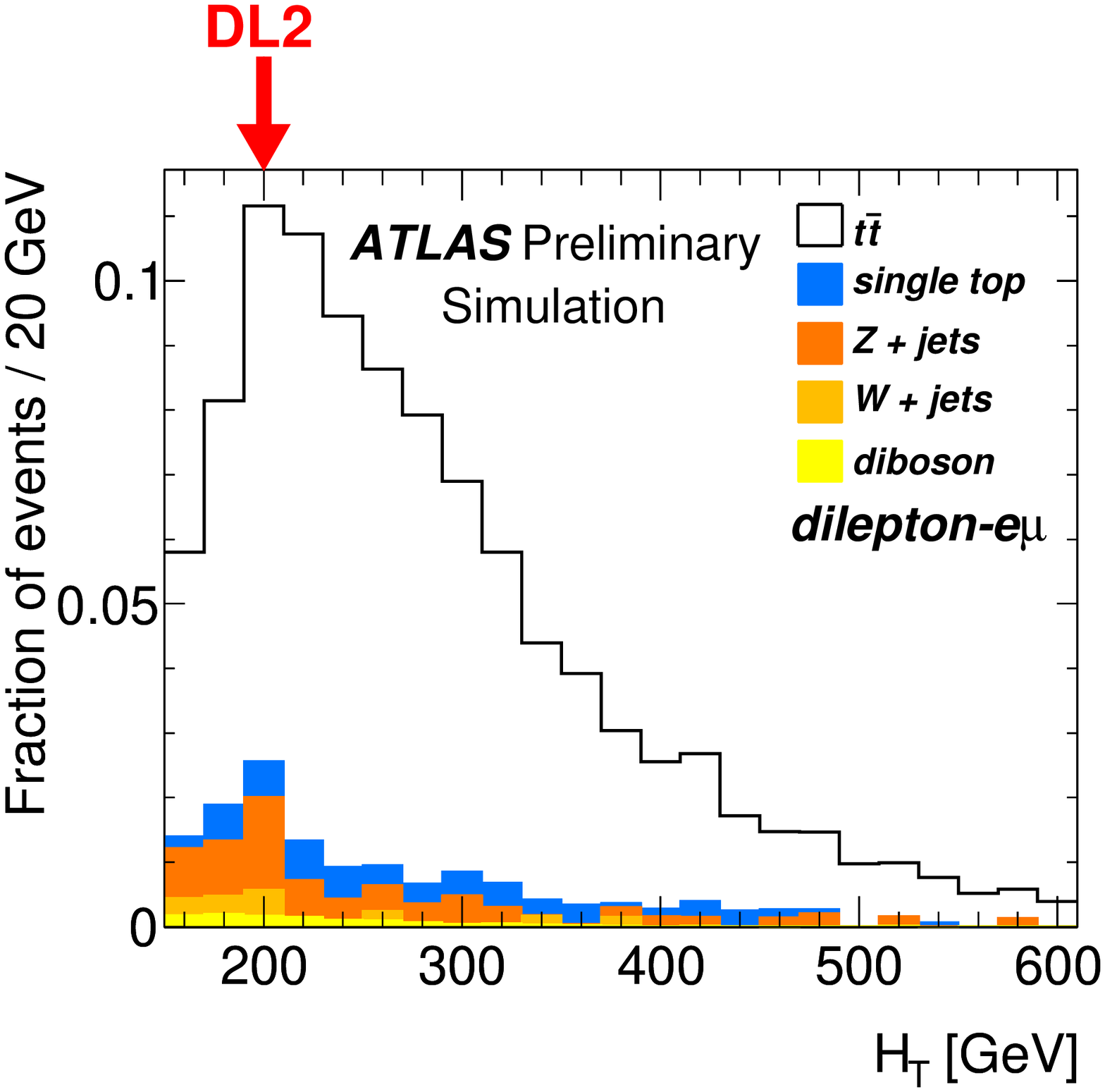}
}
\caption{Left (Right): Distribution of the \met\ ($H_T$) for events passing 
the $ee$ ($e\mu$) dilepton event selection.} \label{f:metht}
\end{figure*}

\begin{figure*}[t]
\centering{
\includegraphics[width=0.495\textwidth]{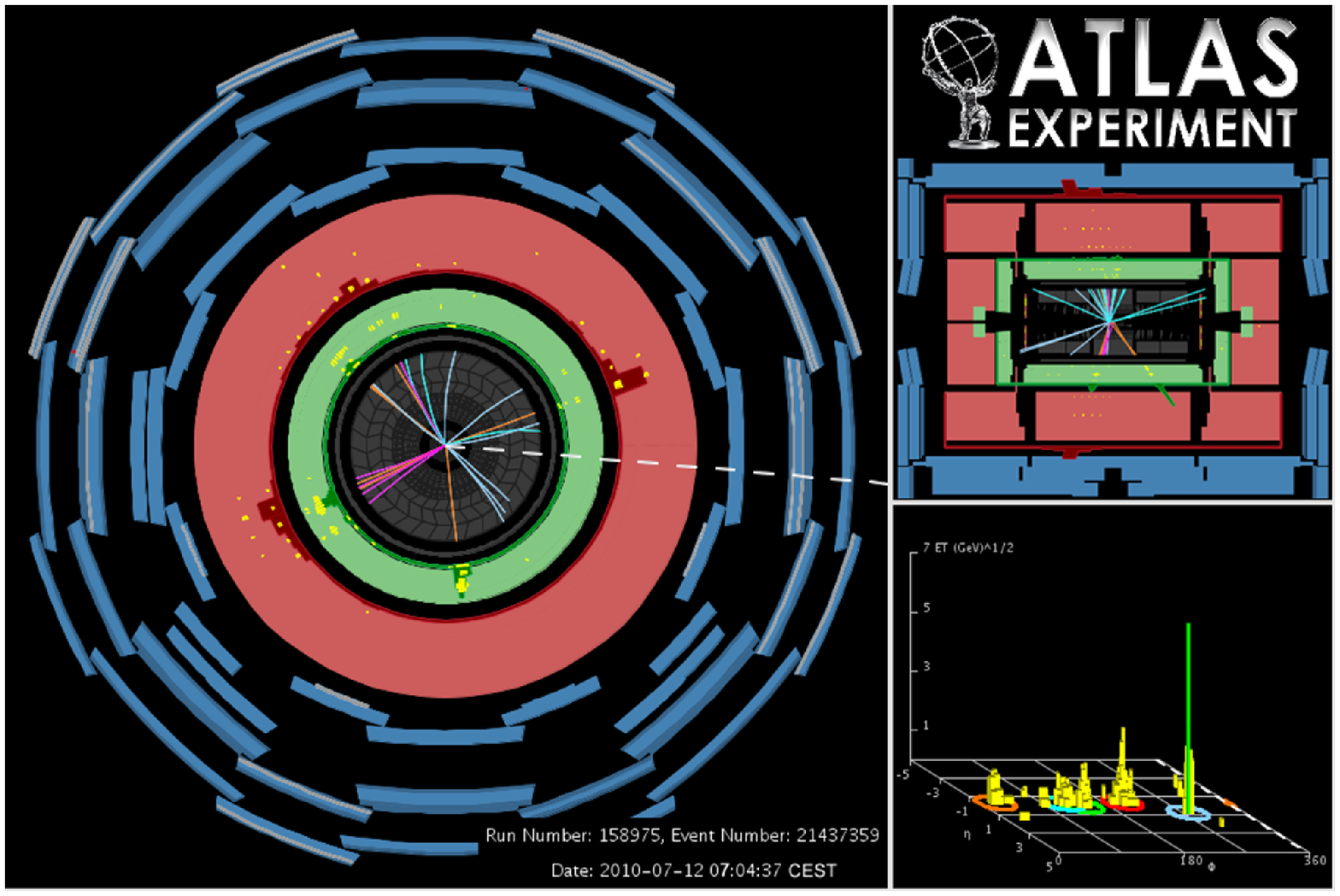}
\includegraphics[width=0.495\textwidth]{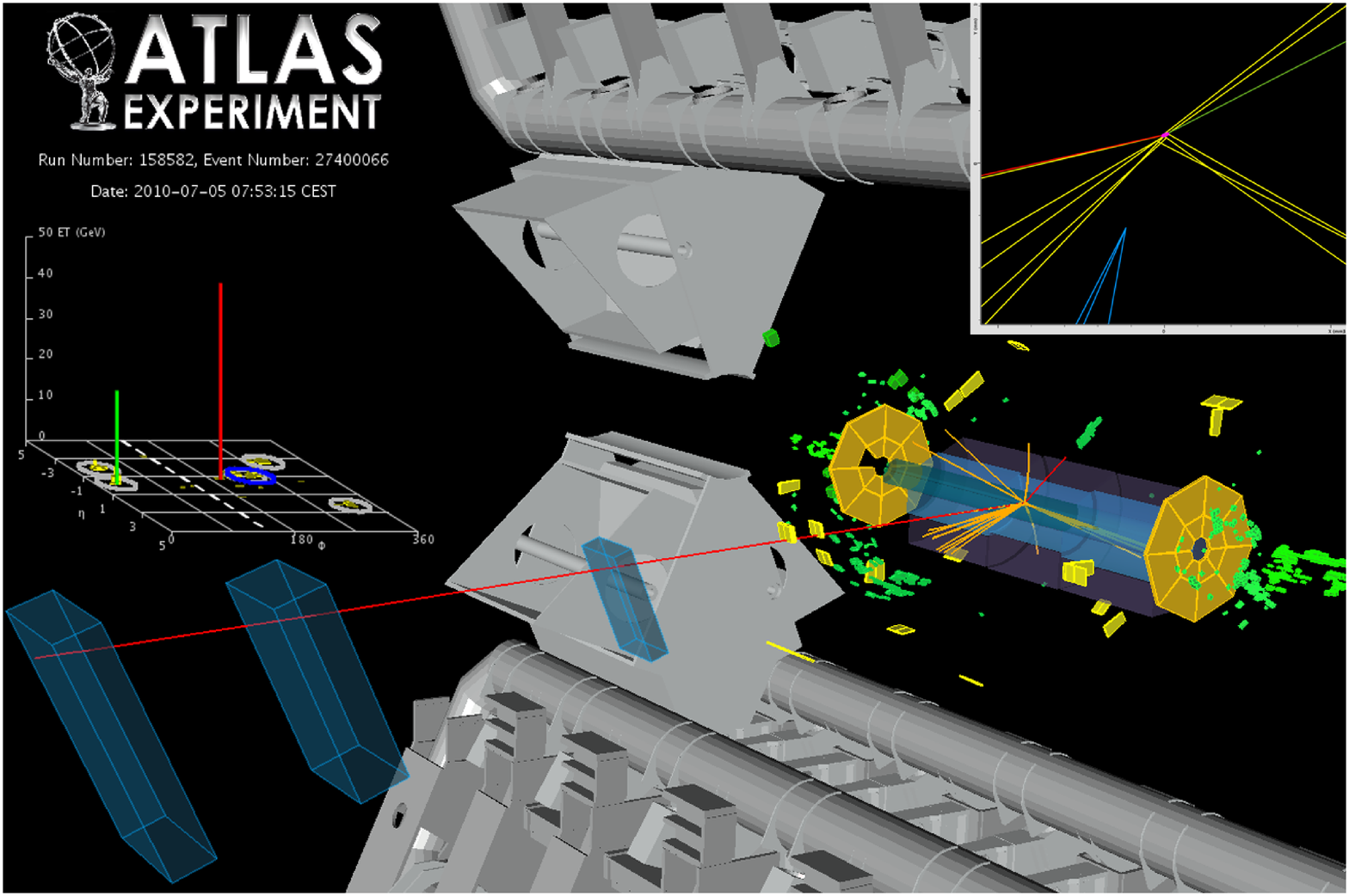}
}
\caption{Event displays for selected candidate events. Shown are on the left an e+jets 
candidate and  on the right a $e\mu$ dilepton candidate. See text for details.} \label{f:evts}
\end{figure*}

A search has been performed for events consistent with top-quark pair
production in \lint\ of 7~\TeV\ $pp$ collision data recorded by ATLAS between
the end of March and mid-July 2010~\cite{ref:topcand}. 

%Several candidate events have observed, in both the lepton plus
%jets and dilepton topologies.

The selection of both lepton plus jets and dilepton \ttbar\ candidates
starts from events collected by single lepton (e/$\mu$, $\pt > 10\GeV$)
level-one triggers.  Events must have a reconstructed primary vertex
with at least 5 tracks, and are discarded if they contain any jet
consistent with out-of-time activity or calorimeter noise.

The selection of lepton plus jets \ttbar events requires the presence
of exactly one offline-reconstructed electron or muon with $\pt > 20
\GeV$. At least four jets with $\pt > 20 \GeV$ and $|\eta| < 2.5$
are then required, at least one of which must be $b$-tagged. Finally, the
\met\ is required to be above 20~\GeV. For background studies in data 
control samples, the event selection is modified to accept events with
one or more jets, with and without \btag requirements.

The selection of dilepton \ttbar\ candidates shares the same baseline
object definition as for the lepton plus jets channel, however, the
requirements on the number of selected jets is relaxed to at least
two, and no \btag is required. Two oppositely-charged leptons ($ee$,
$\mu\mu$, or $e\mu$) each satisfying $\pt > 20 \GeV$ are required. In
the $ee$ channel, to suppress backgrounds from Drell-Yan and QCD
multi-jet events, the missing transverse energy must satisfy $\met >
40 \GeV$, and the dilepton invariant mass must be at least 5 ~\GeV\
away from the \Zboson\ boson mass, i.e. $|m_{ee} - m_{Z}| > 5 \GeV$. For
the di-muon channel, the corresponding requirements are $\met > 30
\GeV$ and $|m_{\mu\mu} - m_Z| > 10 \GeV$. For the $e\mu$ channel,
where the background from $Z \to ee$ and $Z \to \mu\mu$ is expected to
be much smaller, no \met\ or Z-mass veto cuts are applied, but the
event $H_T$, defined as the scalar sum of the transverse energies of
the two leptons and all the selected jets, must be above 150~\GeV.

The distributions of the invariant mass of the 3-jet combination
having the highest \pt, $m_{jjj}$, for events passing the electron
(muon) plus jets selection is reported on the left (right) of
Fig.~\ref{f:mjjj}. The total histogram is normalized to unit area, and
the relative signal and background contributions are from Monte Carlo
expectations.  Similarly, Fig.~\ref{f:metht} reports
the expected \met\ and $H_T$ distributions for events surviving the
$ee$ and $e\mu$ dilepton selections, respectively. In both figures,
the red arrows indicate the corresponding values of the
selected top candidates in the data. In the \lint\ dataset, no
$\mu\mu$ dilepton candidate has been observed.

Selected ATLAS event displays for candidates in the lepton plus jets
and dilepton channels are provided in Fig.~\ref{f:evts}.  An electron
plus jets event is shown on the left: the electron ($\pt=41\GeV$) is
depicted as the orange downward-pointing track associated to the green
cluster, and as the green tower in the $\eta-\phi$ plane lego
plot. The direction of the missing transverse energy ($\met=89\GeV$)
is shown as the dotted line in the $r-\phi$ view. The event contains
four jets with $\pt>20\GeV$, and has an $m_{jjj}$ of
106~\GeV. Fig.~\ref{f:evts} displays on the right an $e\mu$ dilepton
candidate.  The isolated muon ($\pt =48\GeV$) is shown in red, the
isolated electron ($\pt = 23\GeV$) is reported as a red track pointing
to a green cluster. In the $\eta -\phi$ plane lego plot, the
$b$-tagged jet is marked as a blue circle, while the direction of the
missing transverse energy ($\met=77\GeV$) is shown as a white dashed
line.  Finally, the zoom into the primary vertex region allows to
appreciate three displaced blue tracks associated to a secondary
vertex tagged jet. The event has $H_T=196~\GeV$.

\section{Background studies}
\label{sec:bkg}

\begin{figure*}[t]
\centering
\includegraphics[width=0.32\textwidth]{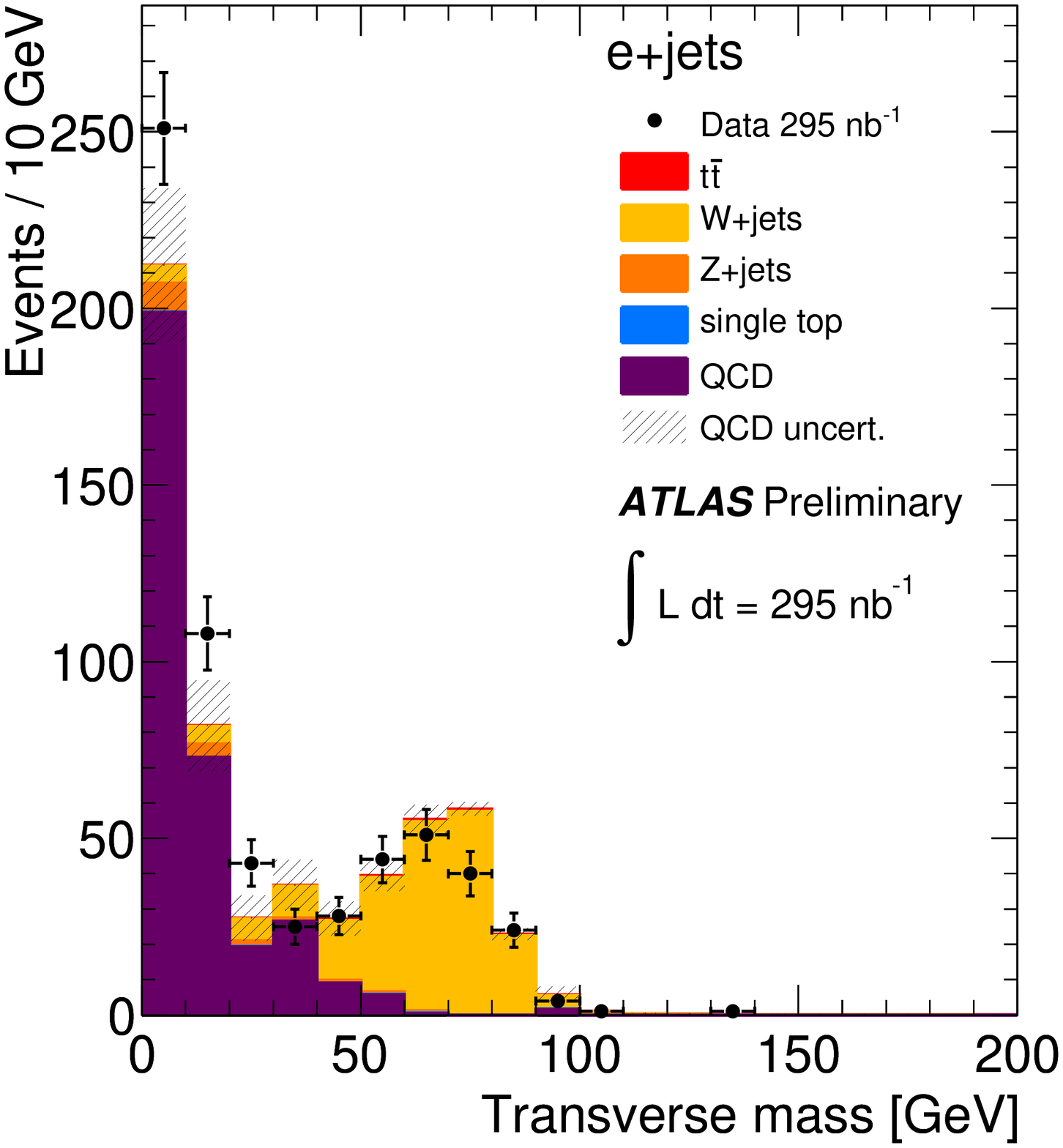}
\includegraphics[width=0.32\textwidth]{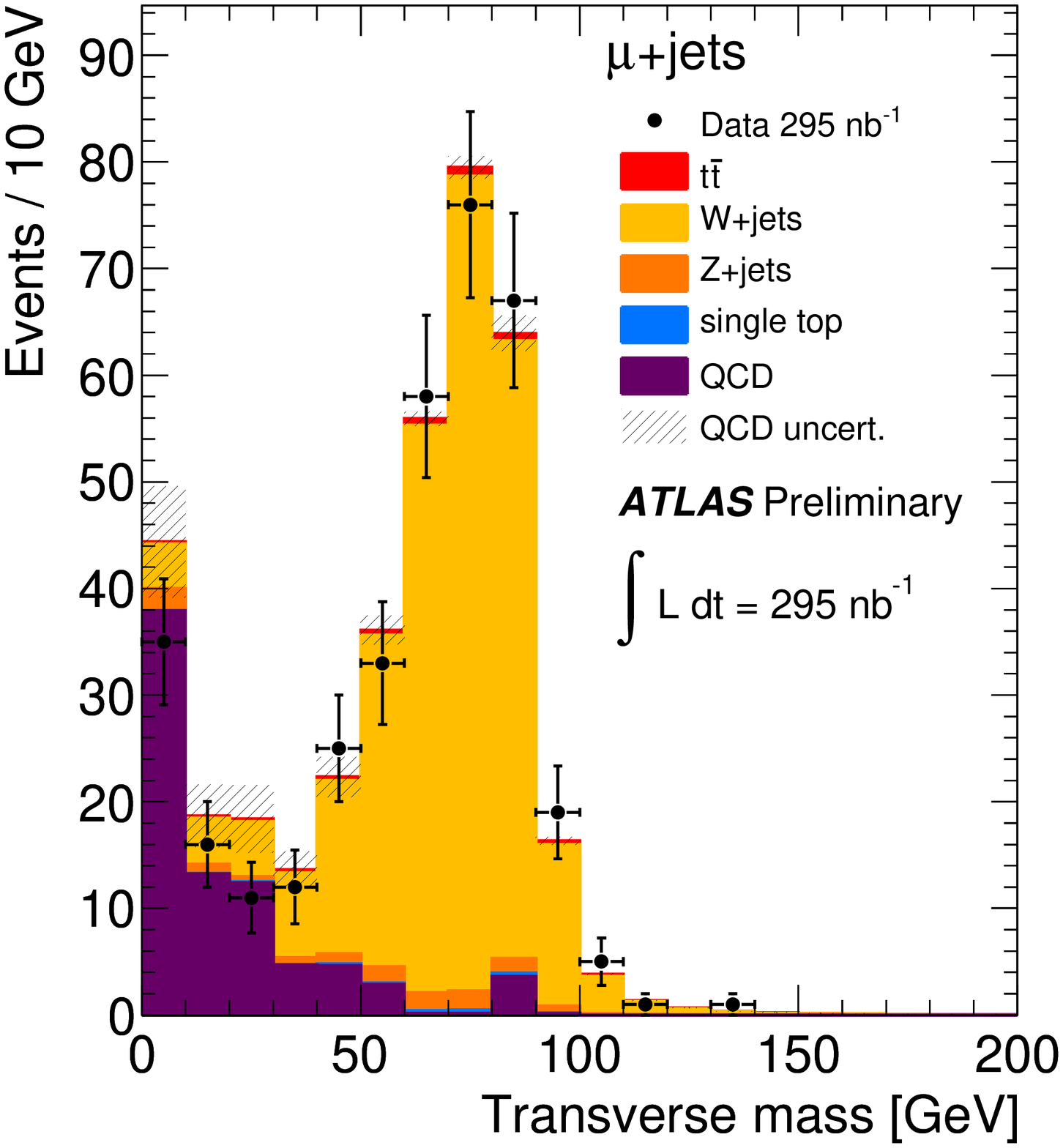}
\caption{Distribution of the transverse mass of the \Wboson\ boson for the 
modified lepton plus jets selection requiring at least one jet with
$\pt>20~\GeV$. The data shown by the points with error bars are
compared to the sum of all expected contributions, taken from Monte
Carlo simulation or estimated using a data-driven technique (QCD
multi-jet). The hatched area shows the uncertainty on the total
expectation due to the statistical error on the QCD background
estimate.}
\label{f:f1}
\end{figure*}

\begin{figure*}[t]
\centering
\includegraphics[width=0.36\textwidth]{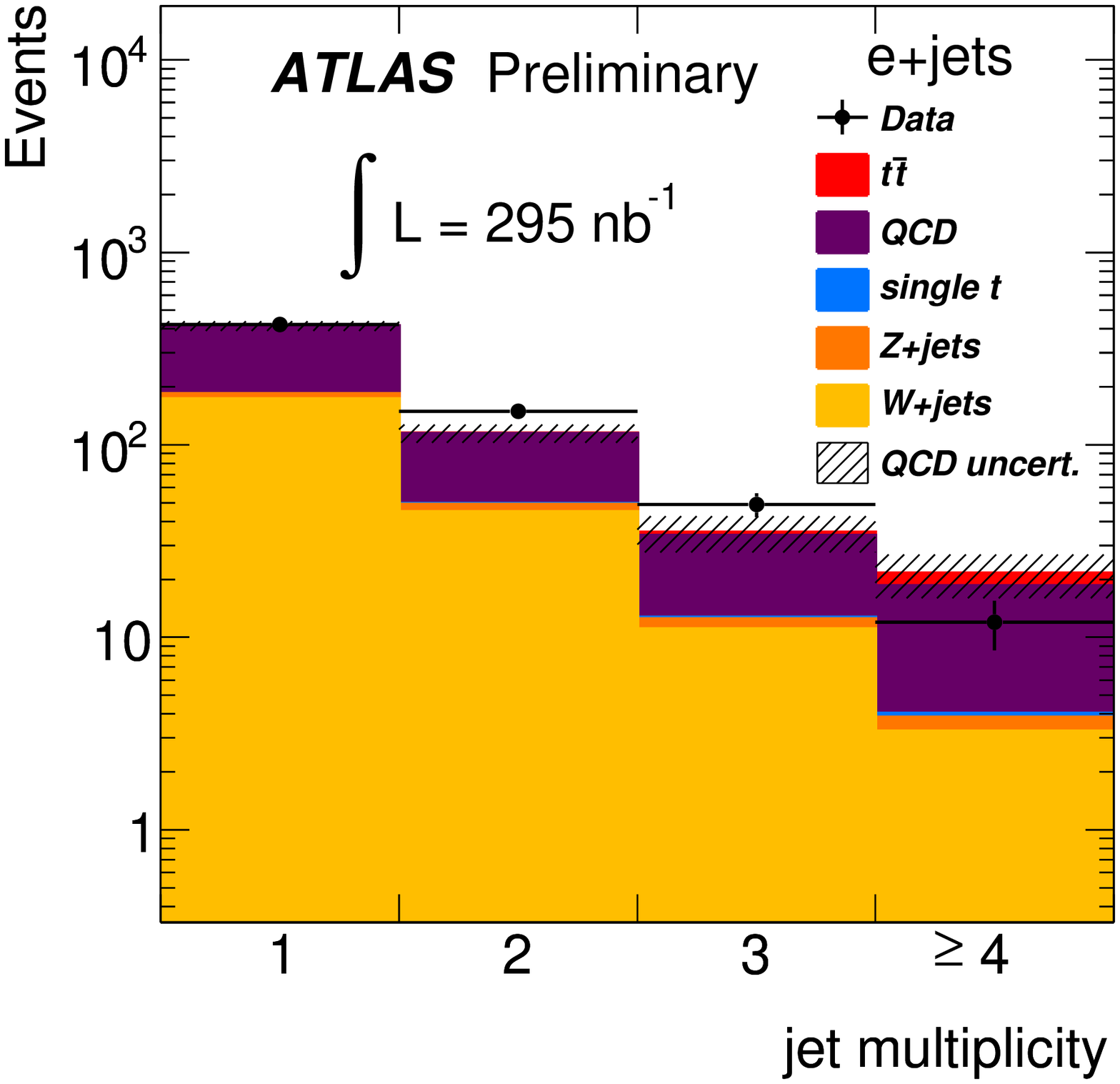}
\includegraphics[width=0.36\textwidth]{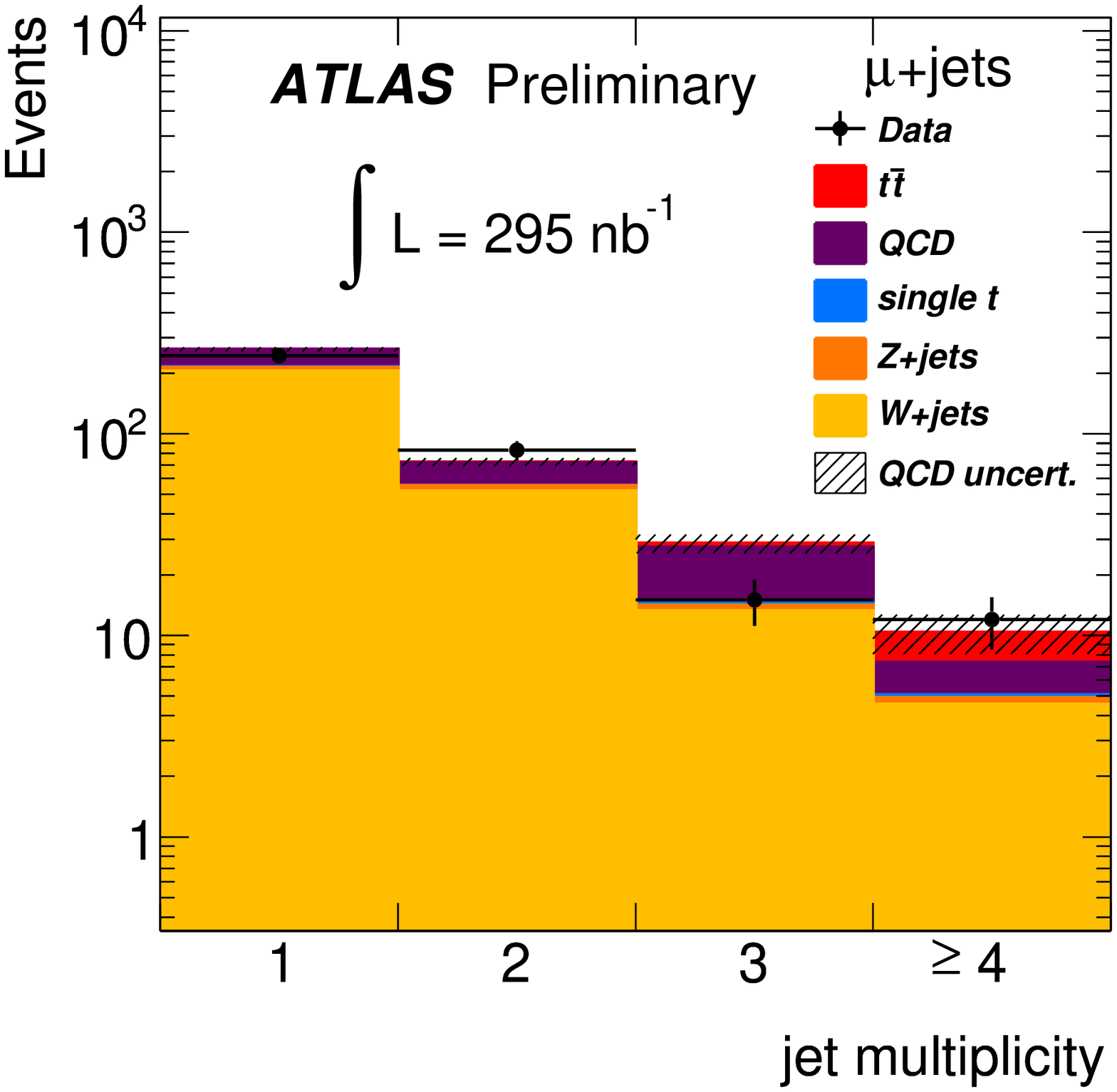}
\includegraphics[width=0.36\textwidth]{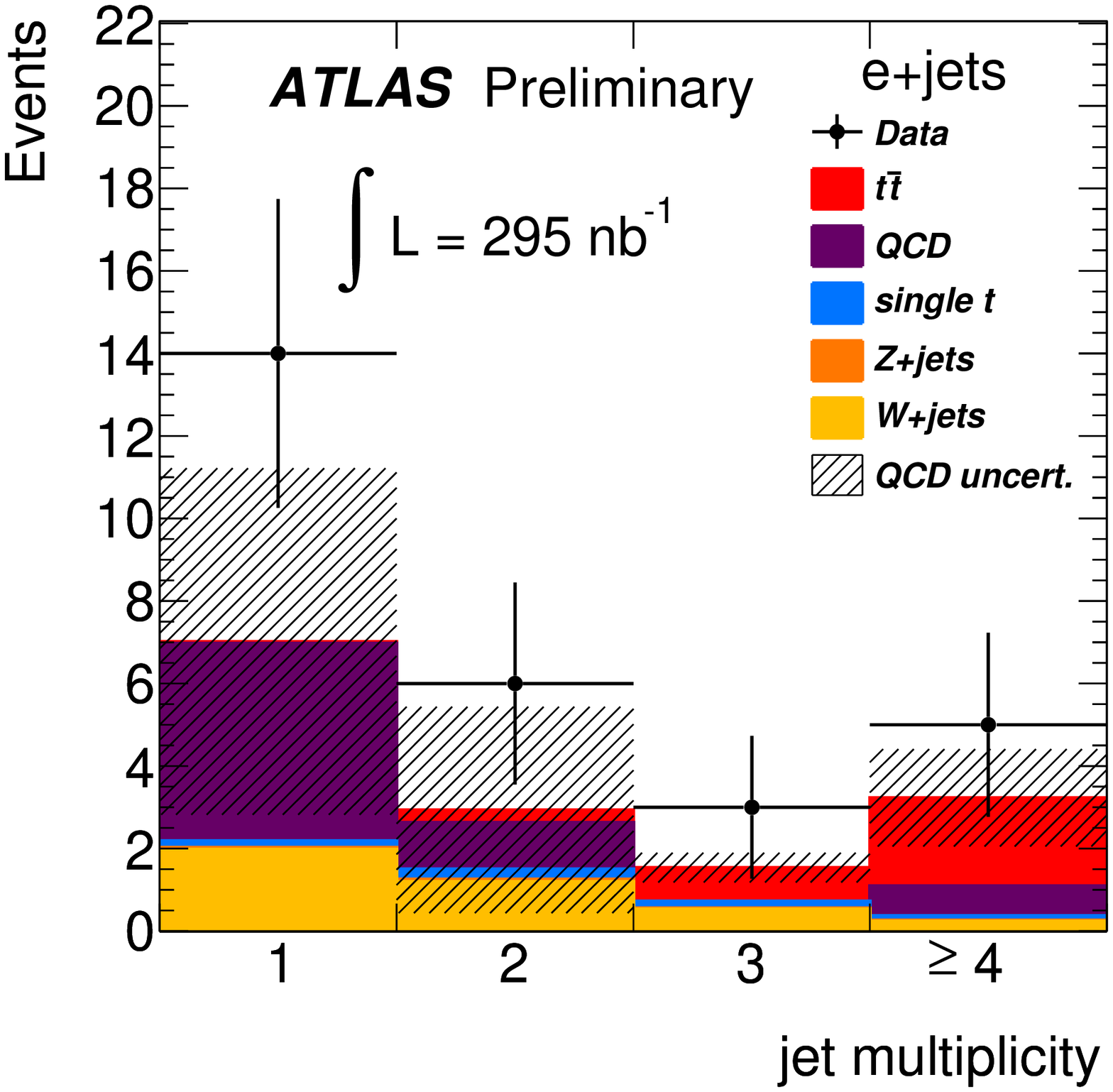}
\includegraphics[width=0.36\textwidth]{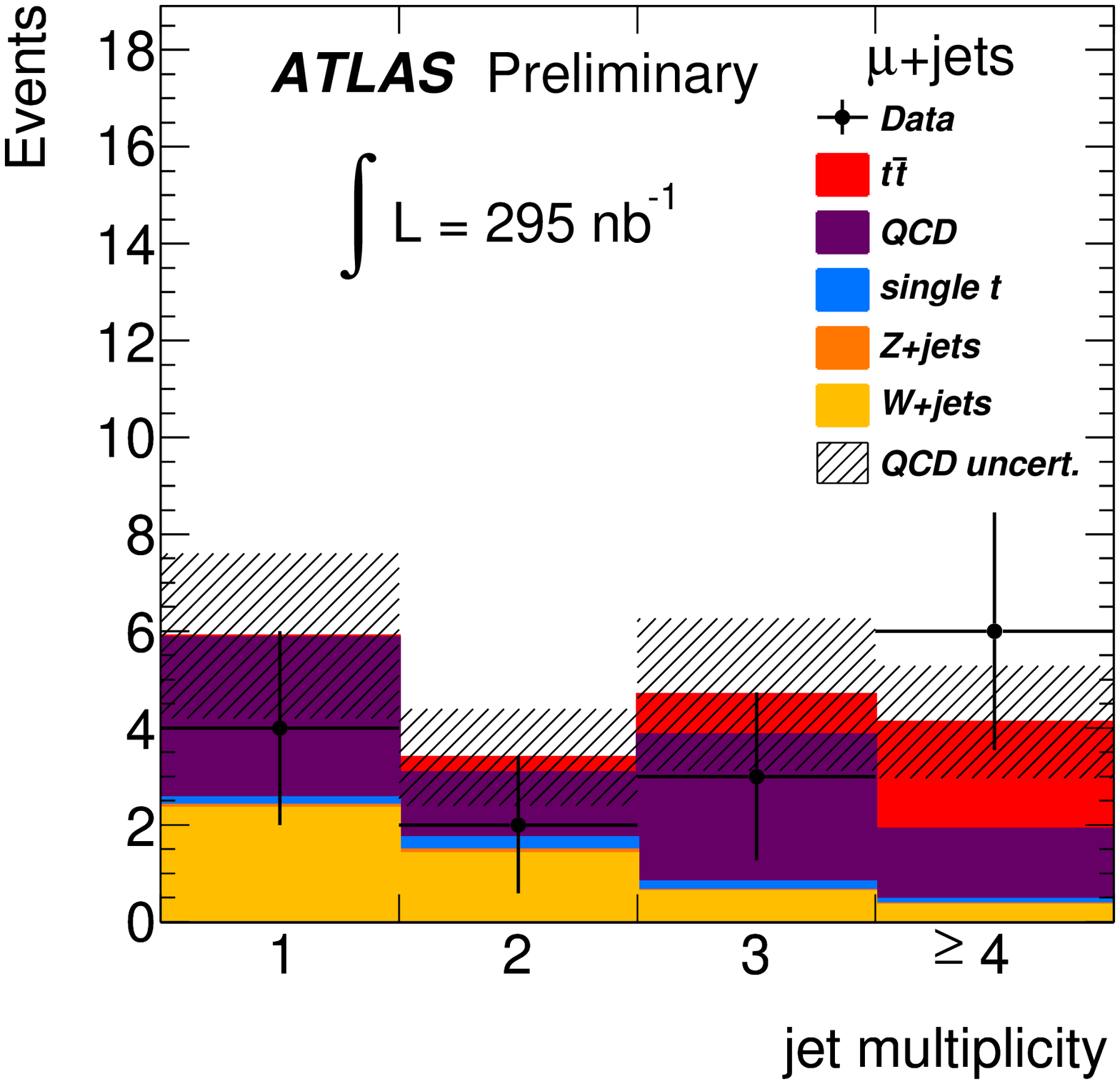}
\caption{Jet multiplicity distributions for the modified lepton plus jets
 selection without (top) and with (bottom) \btag\ requirements.}
 \label{f:f2}
\end{figure*}

In parallel to the selection of top-like events, careful studies of
the background contamination in the selected samples constitute an
essential step in measuring the top pair production rate in ATLAS.
The lepton plus jets channel in particular suffers from significant
background contributions from other Standard Model processes such as
the production of \Wboson\ bosons in association with multiple jets
(\Wj), and QCD multi-jet events. In general these background
contributions are difficult to model reliably using Monte Carlo
information alone, so that data-driven techniques for their estimate
are preferable. The \Wj\ background to the signal region ($N_{jet}\ge
4$) can be constrained by using lower jet multiplicity events, with
and without \btag\ requirements.  On the other hand the QCD multi-jet
background can be studied using background-enhanced control samples,
in low jet multiplicity bins and low \met\ regions.

Currently the data-driven QCD estimate in the lepton plus jet channel
is obtained from the matrix-method technique~\cite{ref:topbkg}.  This
approach is based on the definition of two data regions characterized
by a tight and a loose lepton definition. The data surviving the loose
and tight lepton selections are interpreted as the sum of the
contributions from events with real (\Wboson-like) and fake leptons.
By knowing the efficiencies for a real ($\epsilon_{real}$) and fake
($\epsilon_{fake}$) loose lepton to fulfill the tight selection, the
number of \Wboson-like (including \ttbar) and QCD events can be
derived.  In the current implementation the tight lepton definitions
follow the object requirements described above, while the loose
definitions drop the b-layer hit or the isolation requirements for
electrons and muons respectively.  The $\epsilon_{real}$ is derived
from simulated \Zboson\ decays, while $\epsilon_{fake}$ is determined
from a QCD enhanced sample obtained by requiring the presence of at
least one jet above 20~\GeV\ and $\met <10~\GeV$. To account for
possible \Wboson-like contaminations in the control sample an
iterative correction procedure is applied.

The transverse \Wboson\ boson mass in events fulfilling a modified
lepton plus jets selection, requiring at least one jet with
$\pt>20\GeV$ and no \btag\ is shown in Fig.~\ref{f:f1} for e+jets
and $\mu$+jets events, respectively. The QCD contribution is obtained
from the matrix-method described above. All other contributions are
taken from Monte Carlo expectations. The hatched area shows the
uncertainty on the total expectation due to the statistical
uncertainty on the QCD background estimate. The good agreement between
data and Monte Carlo provides confidence in the ability to correctly
predict both the normalization and the shape of the QCD multi-jet
background with the matrix-method.

The jet multiplicity distributions for events satisfying the modified
selections are shown in Fig.~\ref{f:f2}, before (top row) and after
(bottom row) \btag.  The un-tagged distributions show that
significant samples of $W$+1,2 jets events are becoming available, to
allow the simulation predictions in the $\ge 4$-jet bin to be
constrained with the data. Overall, the data to Monte Carlo agreement
is remarkable also after \btag\ (bottom row), when, despite the low statistics,
the high jet multiplicity data starts to reveal contributions consistent with
\ttbar production.

\section{Conclusions}

All ingredients needed for top-physics analysis are currently
available in ATLAS: reconstruction of leptons, jets and \met\ as well
as b-tagging tools are in a well advanced commissioning stage. In all
aspects, current performances studies, based on early $pp$ collisions
at \rts=7~\TeV, reveal an overall good data to Monte Carlo agreement,
reflecting the maturity of the detector understanding, and the
readiness of the ATLAS experiment to reconstruct complex final states as
those stemming from
\ttbar\ production.

A search for first top candidates in \lint\, has been reported, and
several events with kinematic properties consistent with the \ttbar
production, in both the lepton plus jets and dilepton topologies have
been observed and investigated.
In the lepton plus jet channel preliminary data-driven techniques
have been applied for the determination of the QCD multi-jet
contribution to the selected data sample, which in addition contains
significant numbers of $W$+1,2 jets events that will shortly allow
the Monte Carlo predictions for \Wj\ background to be constrained by
the data.

Although larger data samples will be required to quantify backgrounds to
a level that can support a conclusive top quark observation in ATLAS,
current results are very encouraging and demonstrate the readiness
of the ATLAS collaboration to conduct its top-quark physics program.

%\begin{acknowledgments}
%The authors wish to thank JACoW for their guidance in preparing
%this template.

%Work supported by Department of Energy contract DE-AC02-76SF00515.
%\end{acknowledgments}

%\begin{thebibliography}{9}   % Use for  1-9  references

\end{document}